\documentclass[
    article,
    reprint,
    twocolumn,
    aps,
    prl,
    amsmath,amssymb,
    ]{revtex4-1}
\usepackage{graphicx}
\usepackage{dcolumn}
\usepackage{siunitx}
\usepackage{float}
\usepackage{bm}
\usepackage{amsmath}
\usepackage{mathtools}
\usepackage{afterpage}

\usepackage[utf8]{inputenc}
\usepackage{bm}
\usepackage{soul}
\usepackage{color}
\usepackage{lipsum}
\usepackage{booktabs}
\usepackage{multirow}
\begin{document}

\title{Spin-wave diode and circulator based on unidirectional coupling}

\author{K.~Szulc$^{1}$ }
\author{P.~Graczyk$^{2}$ }
\author{M.~Mruczkiewicz$^{3}$ }
\author{G.~Gubbiotti$^{4}$ }
\author{M.~Krawczyk$^{1}$ }
\email{krawczyk@amu.edu.pl}
\affiliation{%
$^{1}$Faculty of Physics, Adam Mickiewicz University in Poznan, Uniwersytetu Pozna\'{n}skiego 2, 61-614 Pozna\'{n}, Poland 
}%
\affiliation{%
$^{2}$Institute of Molecular Physics, Polish Academy of Sciences, M. Smoluchowskiego 17, 60-179 Pozna\'{n}, Poland
}%
\affiliation{%
$^{3}$Centre of Excellence for Advanced Materials Application (CEMEA), Slovak Academy of Sciences, D\'{u}bravsk\'{a} cesta 9, 845 11 Bratislava, Slovakia 
}%
\affiliation{%
$^{4}$Istituto Officina dei Materiali del CNR (CNR-IOM), Sede Secondaria di Perugia, c/o Dipartimento di Fisica e Geologia, Universit\`{a} di Perugia, I-06123 Perugia, Italy
}%

\begin{abstract}

In magnonics, an emerging branch of wave physics characterized by low-energy consumption, it is highly desirable to realize circuit elements within the scope of spin-wave computing. Here, based on numerical simulations, we demonstrate the functionality of the spin-wave diode and the circulator to steer and manipulate spin waves over a wide range of frequency in the GHz regime. They take advantage of the unidirectional coupling induced by the interfacial Dzyaloshinskii-Moriya interaction to transfer the spin wave between thin ferromagnetic layers in only one direction of propagation. Using the multilayered structure consisting of Py and Co in direct contact with Pt, we obtain sub-micrometer-size devices of high efficiency. In the diode, the power loss ratio between forward and reverse direction reaches 22 dB, while in the four-port circulator, the efficiency exceeds 13 dB. Thus, our work contributes to the emerging branch of energy-efficient magnonic logic devices, where, thanks to short wavelength of spin waves, it is possible to realize nanoscale devices.

\end{abstract}


\maketitle

\section{Introduction}

A diode and a circulator are electronic and microwave components which found wide applications in many devices for signal processing. A diode allows the flow of signal in only one direction, and for microwaves, it is also known as an isolator. It has already equivalents in optics \cite{1995optdiode}, heat transfer \cite{2004heatdiode,2006heatdiode}, acoustics \cite{2009acdiode,2010acdiode}, and spin-Seebeck effect \cite{2014ssdiode}. Diodes for spin waves (SWs) relying on the dipolar \cite{wu2012nonreciprocal,shichi2015spin, grassi2019slow} or interfacial Dzyaloshinskii-Moriya interaction (iDMI) \cite{2015spindiode} were recently proposed. In circulators, the signal going from one port is always directed only to the nearest port, according to the same sense of rotation. It usually consists of three or four ports. Apart from microwaves and photonics, where the circulators have found applications \cite{Schlo_88,Dotsch:05,HARRIS20092035,Smigaj:10}, they have been recently demonstrated also for acoustic waves \cite{Fleury516}, while a demonstration for SWs is still missing. Circulators used in the industry are mostly macroscopic devices. Their miniaturization with the possibility of implementation to real-life systems is a crucial point of the present studies.

Antisymmetric exchange interaction was proposed by \citet{dzyaloshinsky} and \citet{moriya} about 60 years ago. Recently, it has found interest due to induced chirality of the magnetization configuration \cite{Yu2010,Chen13} and nonreciprocity in the SW propagation \cite{udvardi2009,moon2013,cortes2013,Stashkevich2015,cho2015thickness,Garst_2017}. The DMI can exist in bulk noncentrosymmetric crystals \cite{Muhlbauer915} or at the interface between ferromagnetic and heavy metal layers (iDMI). The iDMI is of high interest due to larger DMI parameter value \cite{tacchi2017thickness,samardak2018enhanced}, flexibility in shaping its strength, and the possibility of working at the nanoscale.

In this paper, we propose a layered sequence of ultrathin ferromagnetic films where the presence of iDMI interaction over one layer only, leads to asymmetric or even unidirectional coupling of SWs between the layers. Interestingly, we found that the multilayer composition can work as a SW diode or a three- or four-port SW circulator, in dependence on the particular structurization. The proposed SW diode, based on Py (Ni\textsubscript{80}Fe\textsubscript{20}) and Co ultra-thin films, offers isolation of SW signal in the reverse direction reaching 22 dB with respect to the transmission in the forward direction. Importantly, from the application point of view, the functionality of the device is preserved for a broad GHz-frequency range. We investigate the coupling between SWs in heterogeneous ultra-thin bilayer by numerical frequency-domain and time-dependent simulations. Then we discuss the coupling strength and the SW transmission between the layers in the framework of the coupled-mode theory. Finally, we present possible realizations of the SW devices -- the SW diode and the four-port circulator, with in-depth analysis of their efficiency.

\subsection{Model}

\begin{figure}[!t]
    \centering
    \includegraphics{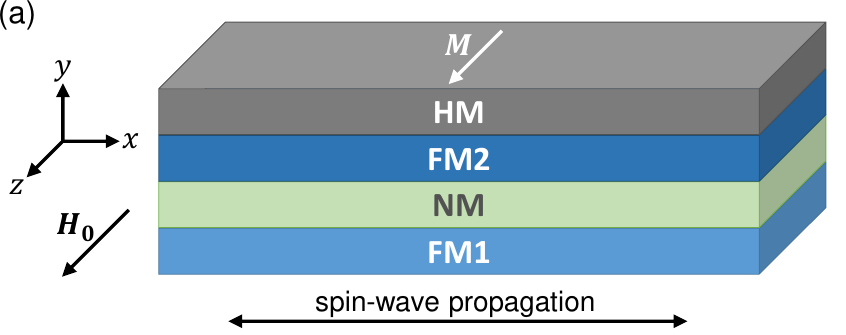}
    \includegraphics{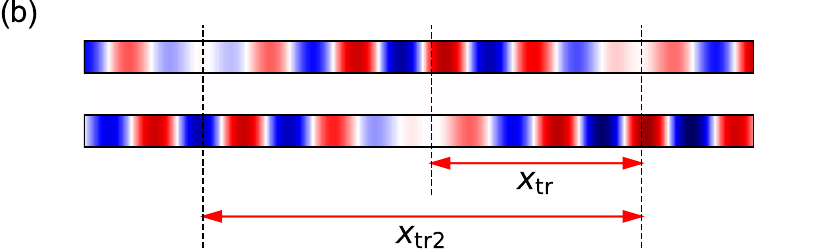}
    \caption{(a) Schematic representation of the multilayer stack and the geometry considered. The layer sequence consists of two ferromagnetic films, FM1 and FM2, separated by the non-magnetic layer (NM). In FM2, the iDMI was induced by the proximity with the heavy metal (HM). Generally, this structure underlies the unidirectional coupling in a wide range of frequency. (b) Transmission length $x_{\mathrm{tr}}$ and $x_{\mathrm{tr}2}$ presented in the FM bilayer system.}
    \label{fig:structure}
\end{figure}

The considered multilayer stack consists of two ferromagnetic (FM) layers separated by a nonmagnetic (NM) spacer, and heavy metal (HM) layer in contact with one of the FM layers [Fig.~\ref{fig:structure}(a)]. We consider SW propagation in the Damon-Eshbach geometry, where the magnetization $M$ and the external magnetic field $H_0$ are aligned in-plane of the films and perpendicular to SW propagation defined by the wavevector $\textbf{k}$.

The Landau-Lifshitz-Gilbert equation for dynamic magnetization components $\textbf{m}$ together with Poisson equation for magnetic scalar potential $\varphi$ are solved by finite-element method in COMSOL Multiphysics in frequency- and time-domain studies in the linear approximation \cite{Graczyk_2018}, i.e., assuming $m_x,m_y \ll m_z \approx M_S$, where $M_S$ is the saturation magnetization. The effective field that exerts a torque on magnetization contains contributions from the Zeeman, exchange, iDMI, and dipolar fields (see Materials and Methods section). We assume that the iDMI is present in the layer FM2 adjacent to the HM, only.

In the first step of calculations we consider multilayer of the Py(3)/NM(5)/Co(2)/Pt composition, where numbers in brackets denote the thickness of the layers in nanometers. For Co layer we assume \mbox{$M_S=956$~kA/m}, exchange stiffness constant \mbox{$A_{\text{ex}}=21$~pJ/m} \cite{moreau2016}, Gilbert damping constant \mbox{$\alpha=0.05$}, iDMI constant \mbox{$D=-0.7$~mJ/m$^2$}, and for Py layer \mbox{$M_S=800$~kA/m}, \mbox{$A_{\text{ex}}=13$~pJ/m}, \mbox{$\alpha=0.005$}, \mbox{$D=0$}. External static magnetic field $H_0$ is fixed to 50 mT.

\subsection{Coupled-mode theory with damping}

The SWs propagating in the system composed of two ferromagnetic layers separated by a nonmagnetic layer are magnetostatically coupled. We can describe this phenomenon using general coupled-mode theory \cite{Yariv1973,Zhang2008} based on the wave properties, only. To describe the interaction between propagating modes, we will use coupling-in-space formalism. The differential equation describing the scalar wave $\psi_l$ propagating in a single layer $l$ is
\begin{equation}
    \frac{\mathrm{d}\psi_l}{dx} = -i \beta_l \psi_l,
\end{equation}
with 
\begin{equation}\label{eq:beta}
    \beta_l = k'_l-i\alpha_l k''_l
\end{equation}
denoting the complex wavevector, where the real part corresponds to the propagation, and the imaginary part to the attenuation of the wave.
For the waves propagating in two coupled layers, we get the mutually dependent differential equations:
\begin{align}
    \frac{\mathrm{d}\psi_1}{dx} = -i \beta_1 \psi_1 + \kappa_{12} \psi_2,\label{eq:couplingdiff1} \\
    \frac{\mathrm{d}\psi_2}{dx} = -i \beta_2 \psi_1 + \kappa_{21} \psi_1, \label{eq:couplingdiff2}
\end{align}
where for the co-directional coupling, i.e., coupling of the waves propagating in the same direction
\begin{equation}\label{eq:couplingpar}
    \kappa_{12} = -\kappa_{21} = \frac{1}{2} \left(\left|k_{\text{P}}-k_{\text{CP}}\right|-\left|k_1-k_2\right|\right)
\end{equation}
are the coupling coefficients. $k_{\text{P}}$ and $k_{\text{CP}}$ are wavevectors of the in-phase and in-counterphase modes of the coupled bilayered system, respectively. Generally, the waves can be described by the complex numbers with the coupling magnitude described with the right side of Eq.~(\ref{eq:couplingpar}). In our case, we are only interested in the magnitude of coupling, and not the phase of the wave, which derives from the argument of $\kappa_{12}$.

The system of differential equations [Eqs.~(\ref{eq:couplingdiff1})~and~(\ref{eq:couplingdiff2})] can be reduced to the homogeneous linear equations. Assuming that the solutions are in the form of $e^{-i\beta x}$, the solvability condition requires that
\begin{equation}
    \beta^2-(\beta_1-\beta_2)\beta+(\beta_1 \beta_2+\kappa_{12}\kappa_{21})=0.
\end{equation}
The solutions of this equation are
\begin{equation}
    \beta_\pm = \bar{\beta} \pm B,
    \label{eq:solutions}
\end{equation}
where
\begin{equation*}
    \bar{\beta} = \frac{\beta_1+\beta_2}{2}, {\;}  B = \sqrt{\Delta \beta^2+\left|\kappa_{12}\right|^2}, \text{ and }
     \Delta \beta = \frac{\beta_1-\beta_2}{2}.
\end{equation*}
Substituting the solutions of the Eq.~(\ref{eq:solutions}) to the Eqs.~(\ref{eq:couplingdiff1})~and~(\ref{eq:couplingdiff2}) and assuming the initial conditions as $\psi_1(0)=A$ and $\psi_2(0)=0$, we end with the general solutions for the coupled wavefunctions
\begin{align}
    \psi_1(x) &= A\left(\cos Bx-i\frac{\Delta \beta}{B}\sin Bx\right)e^{-i\bar{\beta}x}, \label{eq:psi1} \\
    \psi_2(x) &= A\frac{\kappa_{21}}{B}\sin{Bx}\,\, e^{-i\bar{\beta}x}. \label{eq:psi2}
\end{align}
In the synchronous state $k'_1=k'_2=k'$, we can determine transmission length $x_{\text{tr}}$ of the wave from the layer 1 to the layer 2 [see Fig.~\ref{fig:structure}(b)] from zeroing of the term in brackets in Eq.~(\ref{eq:psi1}):
\begin{equation}\label{eq:xtr}
    x_{\text{tr}}=\frac{1}{B}\left(\frac{\pi}{2}-\arctan \frac{i\Delta\beta}{B}\right).
\end{equation}
In the synchronous state, $\Delta \beta=-i (\alpha_1 k''_1-\alpha_2 k''_2)$, so the term in the arctangent is real. In the case when the wave is transferred from the layer with lower damping to the layer with higher damping, the transmission length becomes larger, while in the opposite case, it becomes smaller. If $-\Delta\beta^2>|\kappa_{12}|^2$, then the parameter $B$ becomes imaginary, and if $\alpha_1 k''_1<\alpha_2 k''_2$ then $x_{\text{tr}}<0$ and we can not achieve complete transmission (the structure behaves like an overdamped harmonic oscillator), while if $\alpha_1 k''_1 > \alpha_2 k''_2$ then $x_{\text{tr}}>0$ and we get complete transmission but only once.

We can also extract "there and back transmission" length $x_{\text{tr2}}$ considering the length at which the wave transfers from layer 1 to layer 2 and then transfers back from layer 2 to layer 1 [see Fig.~\ref{fig:structure}(b)]. The solution comes from zeroing of the sine term in Eq.~(\ref{eq:psi2}).
The lowest positive solution is
\begin{equation}\label{eq:xtr2}
    x_{\text{tr2}}=\frac{\pi}{B}.
\end{equation}

At this point, we have to introduce the SW parameters to the coupled-mode theory. Knowing that $\omega' = v_{\text{ph}}k'$ and $\omega'' = v_{\text{gr}}k''$ \cite{gurevich1996magnetization}, the Eq.~(\ref{eq:beta}) is transformed to
\begin{equation}
    \beta_l = \frac{1}{v_{\text{ph},l}}\omega'_l-\frac{i\alpha_l}{v_{\text{gr},l}}\omega''_l,
\end{equation}
where the real $\omega'$ and imaginary $\omega''$ parts of the frequency in the Damon-Eshbach geometry are defined as \cite{gurevich1996magnetization,moon2013}
\begin{widetext}
\begin{equation}
    \omega' = \gamma \mu_0 \left(\sqrt{\left(H_0+\frac{M_S}{4}+\frac{2A_{\text{ex}}}{\mu_0 M_S}k^2\right)\left(H_0+\frac{3M_S}{4}+\frac{2A_{\text{ex}}}{\mu_0 M_S}k^2\right)-\frac{e^{-4|k|d}M_S^2}{16}\left(1+2e^{2|k|d}\right)}+\frac{2D}{\mu_0 M_S}k\right),
\end{equation}
\begin{equation}\label{eq:omegabis}
    \omega'' = \gamma \mu_0 \left(H_0+\frac{M_S}{2}+\frac{2A_{\text{ex}}}{\mu_0 M_S}k^2+\frac{2D}{\mu_0 M_S}k\right).
\end{equation}
\end{widetext}

The value of the $\kappa_{12}$ is determined from the dispersion relation of the coupled bilayer system obtained in the numerical simulations. The parameters in Eq.~(\ref{eq:couplingpar}) are calculated for the given frequency $\omega'$.

\subsection{Coupling parameters}

To describe the coupling between the SWs propagating in a bilayered structure, we define the two coupling parameters between the ferromagnetic layers. The first is the power transfer factor $F_P$, which relies upon the dispersion relation. From the coupled-mode theory, we get that the power transfer factor is \cite{Zhang2008}
\begin{equation}\label{eq:FP}
    F_P = \frac{f_{\text{coup}}^2}{f_{\text{coup}}^2+\Delta f^2},
\end{equation}
where
\begin{align}
    \Delta f &= |f_1-f_2| \text{, and} \\
    f_{\text{coup}} &= f_{\text{P}}-f_{\text{CP}}- \Delta f.
\end{align}

The second parameter is the energy distribution factor $F_E$. We assume that the mode energy of fully-coupled SWs will be shared equally between both ferromagnetic layers. The mode energy for uncoupled SWs will be accumulated in one of the layers only. The total energy density in the $i$-th layer is
\begin{equation}\label{eq:energy}
    E_{i} = E_{i,\text{dip}}+E_{i,\text{ex}},
\end{equation}
where the dipolar energy density $E_{\text{dip}}$ is defined as
\begin{equation}\label{eq:dipolar}
    E_{i,\text{dip}} = \frac{1}{L_i}\frac{1}{2 \mu_0}\iint\limits_{S_i} \textbf{m}\cdot \nabla \varphi \, {\mathrm d}y\, {\mathrm d}x,
\end{equation}
and the exchange energy density $E_{\text{ex}}$
\begin{equation}\label{eq:exchange}
    E_{i,\text{ex}} = \frac{1}{L_i}\frac{A_{\text{ex}}}{M_S^2}\iint\limits_{S_i} (\nabla \textbf{m})^2 \, {\mathrm d}y\, {\mathrm d}x,
\end{equation}
where $L_i$ is the length of the FM layer in the simulations, $S_i = d_i L_i$, where $d_i$ is the thickness of FM layer, and $\textbf{m}=(m_x,m_y)$ is a dynamical component of the magnetization.

The energy distribution factor is defined as follows:
\begin{equation}\label{eq:FE}
    F_E = 1-\frac{1}{2}\left|\frac{E_1^{\text{P}}-E_2^{\text{P}}}{E_1^{\text{P}}+E_2^{\text{P}}}\right|-\frac{1}{2}\left|\frac{E_1^{\text{CP}}-E_2^{\text{CP}}}{E_1^{\text{CP}}+E_2^{\text{CP}}}\right|.
\end{equation}

Values of $F_P$ and $F_E$ are in the range [0,1], where we interpret 0 as no coupling and 1 as a full coupling between SWs propagating in the FM layers.

\section{Results}

\subsection{Unidirectional coupling in the wide range of frequency}

\begin{figure}[!t]
    \centering
    \includegraphics{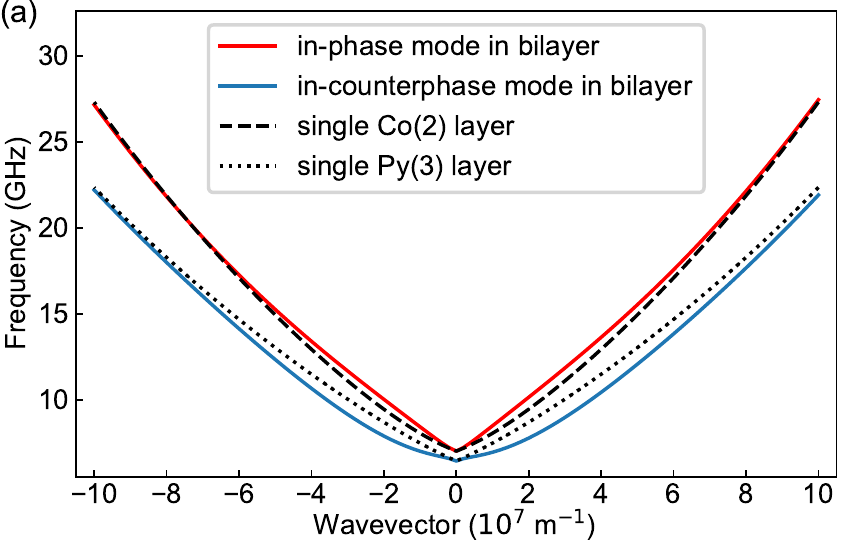}
    \includegraphics{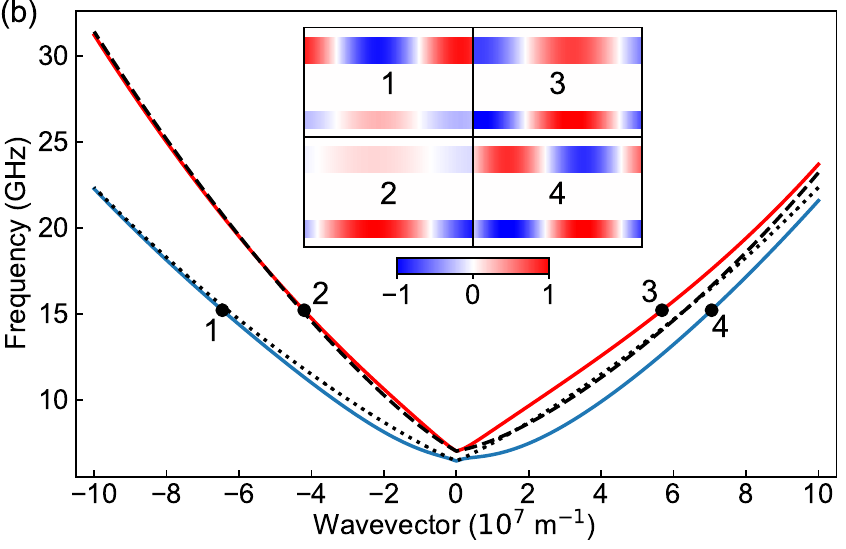}
    \caption{Dispersion relation of SWs as a function of wavevector $k$ in Py(3)/NM(5)/Co(2)/Pt multilayer for iDMI constant in Co layer (a) $D=0$ and (b) $D=-0.7$~mJ/m$^2$. For reference, we show the dispersion relation of SWs in the uncoupled Pt/Co and Py layers with dashed and dotted lines, respectively. In the insets in (b), we show the $m_x$ amplitude of the SWs propagating in both directions at 15.2 GHz. For $D=0$, dispersion relation is almost symmetrical with respect to $k=0$. For $D=-0.7$~mJ/m$^2$, the DMI breaks the symmetry leading to strongly coupled modes in the $+k$ range (insets 3 and 4), and single layer excitation in the $-k$ range (insets 1 and 2). The highest coupling occurs in the region of overlapping of the dispersion relation of the uncoupled layers.}
    \label{fig:dispersion}
\end{figure}

The first step of the investigation of the SW dynamics is the calculation of a dispersion relation. In Fig.~\ref{fig:dispersion} we plotted the dispersion relations of Py(3)/NM(5)/Co(2)/Pt multilayer (solid lines) and uncoupled Co(2)/Pt (dashed lines) and Py(3) (dotted lines) layers for two different values of iDMI constant. For $D=0$ [Fig.~\ref{fig:dispersion}(a)] all dispersion relations are almost symmetric with respect to $k=0$ with only small asymmetry related to dipolar interaction. A small change of the dispersion relation for multilayer, in comparison to the uncoupled layers, is the effect of weak coupling between the FM layers. Taking nonzero iDMI constant, we introduce strong nonreciprocity to the SW dispersion of the mode related to the Co layer. Interestingly, we found that for $D=-0.7$~mJ/m$^2$ [see, Fig.~\ref{fig:dispersion}(b)] the dispersion relation for Co layer almost overlaps with the dispersion relation for Py in the broad range of positive wavevector. Since both modes have almost the same frequency (resonance) and wavevector (phase matching), one can expect strong interaction between them in the multilayer system \cite{Zhang2008}. Two interacting modes are hybridized forming collective excitations, with in-phase [see inset 3 in Fig.~\ref{fig:dispersion}(b)] and in-counterphase [see inset 4 in Fig.~\ref{fig:dispersion}(b)] SW modes at a higher and lower frequency, respectively \cite{Mruczkiewicz2013,Graczyk_2018}. Indeed, we can see the repulsion of the dispersion branches related to the in-phase and in-counterphase modes for positive $k$ in the multilayer (see the red and blue curves in Fig.~\ref{fig:dispersion}(b)), being the effect of strong dipolar coupling between modes in Py and Co. For the negative wavevectors, the dispersions for the uncoupled FM layers are well separated, and in the multilayer, they follow the same lines pointing at the weak coupling between FM layers [see insets 1 and 2 in Fig.~\ref{fig:dispersion}(b)].
Comparing both dispersions in Fig.~\ref{fig:dispersion}(a) and \ref{fig:dispersion}(b) we conclude that adding iDMI to Co layer can lead to strong SW coupling between FM layers for the waves propagating in one ($+k$) direction, while in the structure without iDMI, the coupling can be weak and symmetrical. The general procedure for achieving unidirectional coupling is described in the Materials and Methods section.

\begin{figure}[!t]
    \centering
    \includegraphics{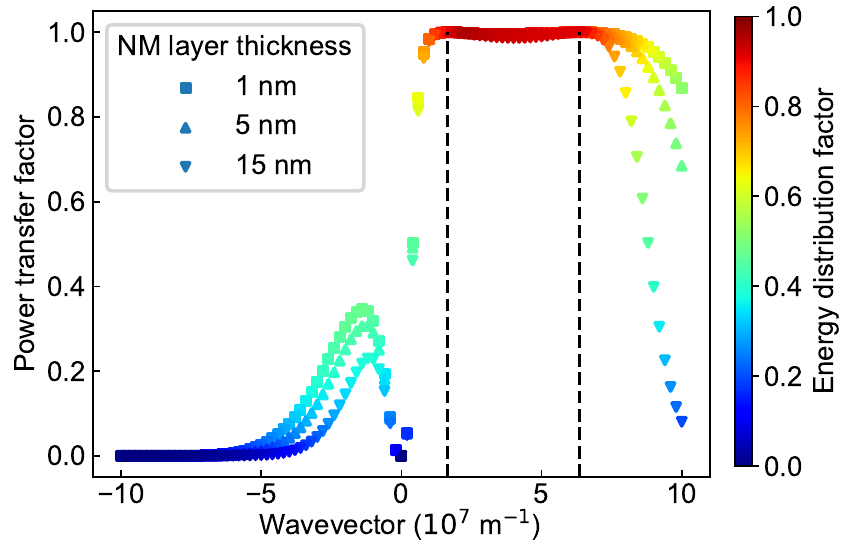}
    \caption{Power transfer factor $F_P$ (the vertical axis) and energy distribution factor $F_E$ (color scale) as a function of the wavevector in Py(3)/NM/Co(2)/Pt multilayer with iDMI constant in Co layer $D=-0.7$~mJ/m$^2$ for three different thicknesses of the NM layer. We reach coupling parameters close to a maximum value of 1 in the range between $1.6\times10^7$ and $6.3\times10^7$ m$^{-1}$ (marked with dashed lines). In the negative $k$ range the coupling is weak and reduces with the increase of the NM layer thickness.}
    \label{fig:coupling}
\end{figure}

For further investigations, the determination of the SW coupling is the crucial point. For this purpose, we use the coupling parameters defined in Eqs. (\ref{eq:FP}) and (\ref{eq:FE}). In Fig.~\ref{fig:coupling} we plot $F_P$ (vertical axis) and $F_E$ (color of the points) in Pt/Co(2)/NM/Py(3) multilayer with $D=-0.7$~mJ/m$^2$ for different thicknesses of NM layer in dependence on the wavevector of the SW. On the positive $k$ side, both coupling parameters are very close to the maximum value in the range between two dispersion crossing points ($1.6\times10^7$ and $6.3\times10^7$ m$^{-1}$). That means the SWs are nearly fully coupled in a wide range of wavevector and frequency. On the negative $k$ side, the coupling is significant only in the long-wavelength range, reaching its maximum for $k\approx -2\times10^7$~m$^{-1}$. The increase of the thickness of the NM spacer leads to a decrease of coupling parameters, except the range of strong coupling between the dispersion crossing points. It is ascribed to the weaker dipolar interaction between the layers.

\begin{figure}[!t]
    \centering
    \includegraphics{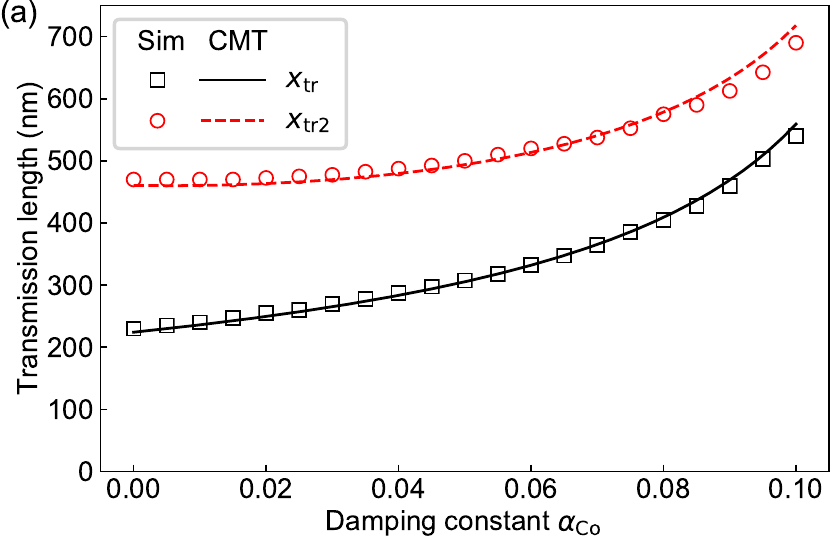}
    \includegraphics{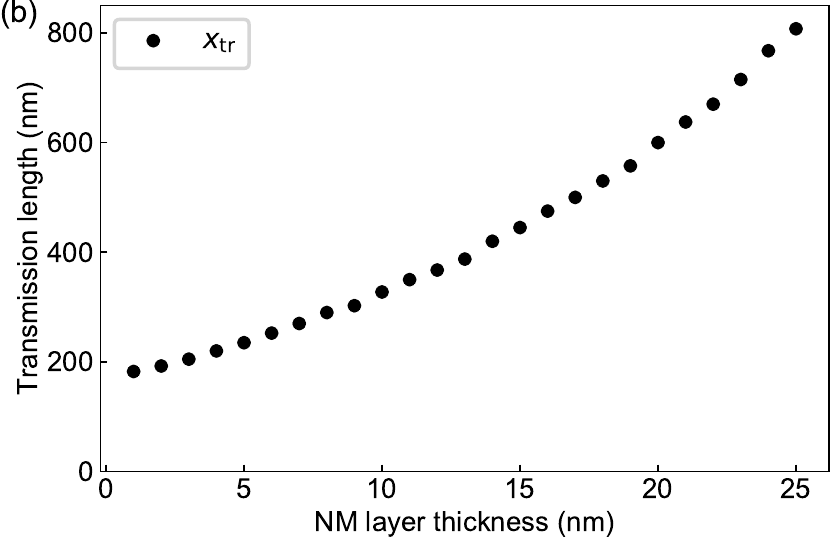}
    \caption{Transmission length dependence on (a) the damping constant in the Co layer in the Py(3)/NM(5)/Co(2)/Pt multilayer with the damping in the Py $\alpha_{\text{Py}}=0.005$ and (b) the thickness of NM layer in the Py(3)/NM(x)/Co(2)/Pt multilayer. The source of the SW of 15.2 GHz is located in the Py layer. In (a), the simulations (Sim) results were compared with the Eqs.~(\ref{eq:xtr})~and~(\ref{eq:xtr2}) from the coupled-mode theory (CMT). In (b), we present the simulations results for $x_{\rm{tr}}$ only.}
    \label{fig:transmission}
\end{figure}

Another important parameter associated with the coupling between the two layers is the transmission length defined in Eqs.~(\ref{eq:xtr}) and (\ref{eq:xtr2}). Many parameters affect this physical quantity. We focused on two of them, which are important in our study -- the damping constant and the NM layer thickness. In Fig.~\ref{fig:transmission}(a), we show the transmission length in the Py(3)/NM(5)/Co(2)/Pt multilayer depending on the damping constant in the Co layer. In the simulations, the SW source emitting the SW at frequency $\omega'/2\pi=15.2$ GHz is located in the Py layer. From Eq.~(\ref{eq:omegabis}) we get $\omega''/2\pi=20.6$ GHz for Co layer and 19.2 GHz for Py layer. Results from the numerical simulations were compared with the Eqs.~(\ref{eq:xtr})~and~(\ref{eq:xtr2}) derived from the coupled-mode theory. We get a satisfying agreement between these approaches. Both $x_{\text{tr}}$ and $x_{\text{tr2}}$ are increasing with the increase of the damping constant. However, $x_{\text{tr}}$ is growing faster than $x_{\text{tr2}}$ leading to the conclusion that the transmission length from the layer with higher damping (Co layer) to the layer with lower damping (Py layer) is decreasing with the increase of the damping constant. In Fig.~\ref{fig:transmission}(b), we show the transmission length in the Py(3)/NM(x)/Co(2)/Pt multilayer depending on the NM layer thickness. We assumed $\alpha_{\text{Co}}=\alpha_{\text{Py}}=0$. The transmission length is increasing with the exponential character of growth. When $\Delta\beta=0$, Eq.~(\ref{eq:xtr}) reduces to the form $x_{\text{tr}}=\pi /(2 |\kappa_{12}|)$. Thus, the coupling coefficient is decreasing exponentially with the increase of the separation between the layers \cite{1981grunberg}.

\subsection{Spin-wave diode}

\begin{figure}[!t]
    \centering
    \includegraphics{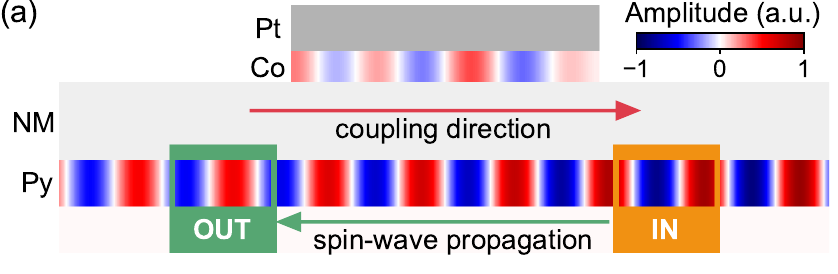}
    \includegraphics{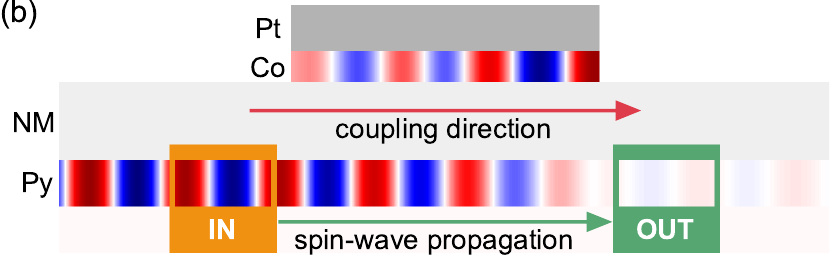}
    \includegraphics{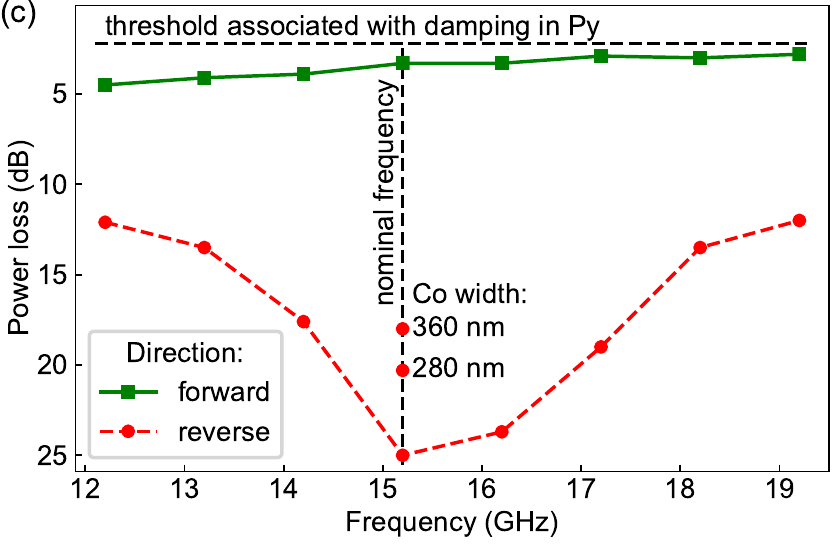}
    \caption{ (a-b) Propagation of the SW in a diode at 15.2 GHz frequency in Py(3)/NM(5)/Co(2)/Pt in (a) forward and (b) reverse direction for the width of the Co/Pt stripe 320 nm. In (a), the SW propagation direction is opposite to the coupling direction, so that the SW transfers weakly to the Co stripe, and we get a signal of high intensity in the output. In (b), the SW propagation direction is the same as the coupling direction, so that the SW transfers to the Co stripe, where is strongly damped, leading to the low intensity of the signal at the output. (c) The power loss in the forward and reverse direction in dependence on the frequency. Additional points at nominal frequency represent power loss for two different widths of Co stripe. In a broad frequency range of 7 GHz, the SW diode preserves its strong isolating properties.}
    \label{fig:diode}
\end{figure}

Taking into account the unidirectional coupling discussed above, we can design the SW diode. The proposed structure is shown in Fig.~\ref{fig:diode}. It consists of continuous Py film, which is the medium where the SWs propagate from the input to the output and Co/Pt stripe, which is a functional element of a diode where iDMI introduces nonreciprocal interaction. They are separated by a 5-nm-thick NM spacer which is sufficient to neglect RKKY interaction. We chose the frequency of the SW from the crossing point of the dispersion relation of uncoupled layers shown in Fig.~\ref{fig:dispersion}(b) to get the full coupling between FM layers. The width of the Co/Pt stripe is matched to the transmission length $x_{\rm{tr}}$, which is related to the coupling strength and the damping in the layers. To determine the efficiency of the device, we calculate the power loss $d_P = 10\log \left(E_{\text{IN}}/E_{\text{OUT}}\right)$, where $E_{\text{IN}}$ is the energy measured in the steady state in front of the device and $E_{\text{OUT}}$ -- behind the device, calculated according to Eq.~(\ref{eq:energy}).

The operation of the diode is depicted in Fig.~\ref{fig:diode}(a)-(b), which shows results from the time-domain simulations of SW continuously excited at the 15.2 GHz frequency in Py at the input (IN). We fixed the width of the Co/Pt stripe to 320 nm. Due to weak coupling between the SWs propagating in the $-x$ direction [Fig.~\ref{fig:diode}(a)], the transmission to the Co stripe is small, and the SW passes the diode retaining its intensity. The total power loss in this direction reaches 3.3 dB, and it is mainly due to the Gilbert damping in Py (2.2 dB). On the other hand, the SW propagating in the $+x$ direction [Fig.~\ref{fig:diode}(b)] transfers almost entirely to the Co stripe where is strongly attenuated due to the high damping. Some residual intensity at the output is the effect of incomplete transfer to Co and return transfer from Co after reflections from the boundaries of the stripe. In fact, along the reverse direction, the total power loss increases to 25 dB. To sum up, the difference in the SW energy in the forward and reverse direction equals 21.7 dB.

We investigated the efficiency of the Py(3)/NM(5)/Co(2)/Pt SW diode in the wide range of frequency. Obtained results of the power loss in both directions of propagation are collected in Fig.~\ref{fig:diode}(c). Although SW transmission length varies in dependence on the frequency, the structure preserves strongly asymmetric transmission. In the forward direction, the diode works as well as in nominal frequency. The power loss in the reverse direction is reduced but remains significantly higher than in the forward direction. Estimated relative frequency range in which the device works is $\Delta \omega'/\omega'_0\approx 0.5$ ($\omega'_0/2\pi = 15.2$ GHz). Additional simulations were made to check the efficiency in the structures with different width of the Co stripe, i.e., for 280 and 360 nm at 15.2 GHz, in reverse direction only [see Fig.~\ref{fig:diode}(c)]. The diode preserves its high isolation in a broad range of the coupler width.

Moreover, we investigated the SW diode working with SWs of longer (390 nm) wavelength, which should simplify the detection of the effect experimentally. We selected another crossing point from Fig.~\ref{fig:dispersion}(b), located at 8.2 GHz frequency. The width of the Co stripe was set to 190 nm. We obtained a power loss of 6.7 dB in the forward direction and 14.6 dB in the reverse direction. In this case, we distinguished three mechanisms responsible for the smaller efficiency of the diode. First, SWs of longer wavelength are coupled stronger than SWs of shorter wavelength. This effect is shown in Fig.~\ref{fig:coupling}. The SW at 8.2 GHz corresponds to $k=1.6\times10^7$ m$^{-1}$. The coupling for negative $k$ reaches its maximum in the vicinity of this point. This effect leads to an additional decrease in the signal in the forward direction. Second, the width of the Co stripe is too small to attenuate the SWs in reverse direction effectively. Third, the SW tends to reflect partially inside the Py layer in the points where the Co layer has its boundaries, which leads to additional losses. Besides these limitations, which can be further optimized, the structure still is efficient enough to be considered as a diode.

\begin{figure*}[!t]
    \centering
    \begin{tabular}{cc}
        \includegraphics{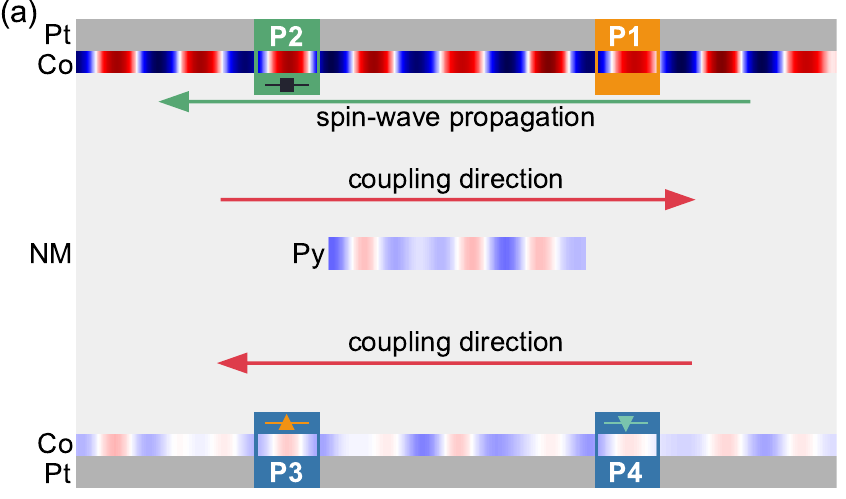} & \includegraphics{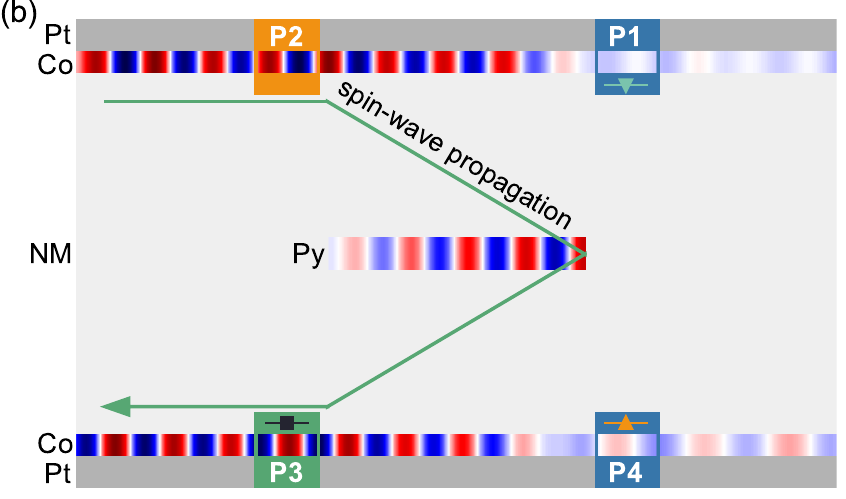} \\
        \includegraphics{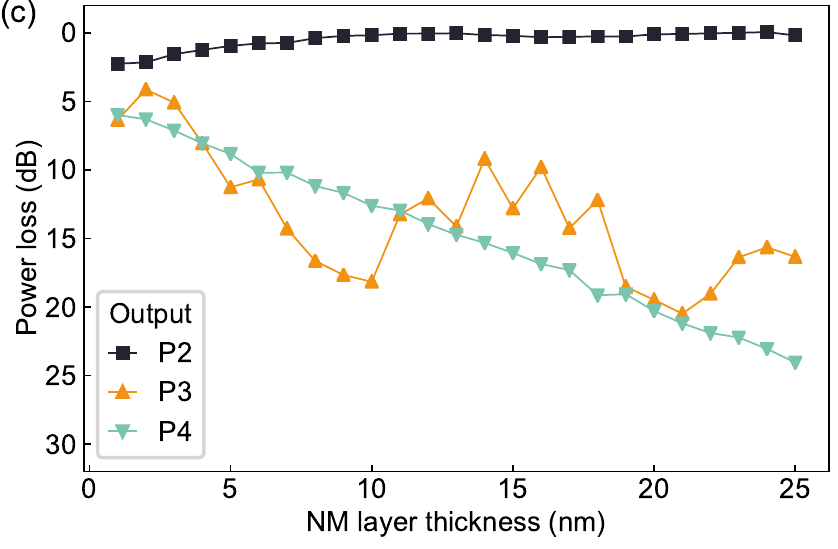} & \includegraphics{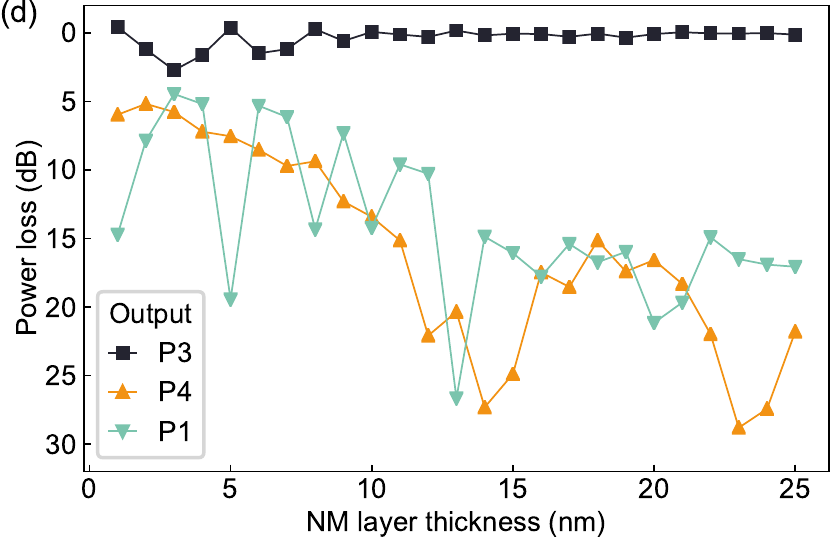}
    \end{tabular}
    \caption{ (a-b) Model of the four-port SW circulator based on the multilayered structure with unidirectional coupling. Propagation of the SW of 15.2 GHz frequency in Pt/Co(2)/NM(15)/Py(3)/NM(15)/Co(2)/Pt circulator in the (a) non-coupling direction with the SW source located in the port P1 and (b) coupling direction with the SW source located in the port P2. In (a), the SW transfers weakly to the Py layer, so it goes mainly to the port P2. In (b), the SW transfers from the upper Co layer to Py, and after the reflection from the right side of Py, it transfers to the lower Co layer, reaching the port P3 in the end. A small-amplitude signal visible at isolated ports is a result of weak coupling between Co layers. (c-d) The power loss measured in the output ports in regard to the input port in dependence on the NM layer thickness for the input located in the (c) port P1 and (d) port P2. In (c), the power loss in the target port P2 is increasing and in the rest of the ports is decreasing with the increase of NM layer thickness. In (d), the power loss in the target port P3 and port P1 is fluctuating due to the resonance in the Py stripe. The minima of the power loss are corresponding to the Py stripe width fulfilling the resonance condition.}
    \label{fig:circulator}
\end{figure*}

\subsection{Spin-wave circulator}

Next, we further exploit the unidirectional coupling to design a SW circulator. The schematic structure of the four-port circulator is shown in Fig.~\ref{fig:circulator}. As compared with the structure of the diode, an additional FM layer is present on the opposite side of the stripe, playing the role of two additional ports. To get the functionality of the circulator, we need the stripe, which is unidirectionally coupled with both top and bottom layers but in opposite directions of SW propagation. We achieved this condition by taking identical outer layers having opposite iDMI constant and the inner stripe lacking iDMI. In our case, we propose to use Py as an iDMI-free coupling stripe and Co/Pt as guiding layers with swapped order in the bottom and top layers. The separation between the stripe and the layer was increased to 15 nm to reduce the dipolar coupling between the Co layers. We keep the width of the Py stripe sufficient to transfer the SW fully from one layer to another, thus for 15.2 GHz, we assume 440 nm. The SWs have to be transmitted from Co to Py as well as from Py to Co, and the coupled-mode theory implies that the maximum efficiency will be achieved when the damping constants in all layers are equal \footnote{If $\alpha_{\text{Co}} \neq \alpha_{\text{Py}}$, then $x_{\text{tr, Co}\rightarrow\text{Py}}\neq x_{\text{tr, Py}\rightarrow\text{Co}}$ so that full transfer can not be achieved.}. Thus, in the following simulations, we assume the damping constant in Co to be $\alpha_{\text{Co}} = 0.01$. Moreover, we perform simulations with assuming no damping to check the efficiency in the ideal case. The structure has a center of symmetry, therefore, the ports on the same diagonal, i.e., P1 and P3 as well as P2 and P4, work identically, and it is sufficient to investigate only two cases -- propagation in the coupling and the non-coupling direction. 

In the non-coupling case, the SW source is located in the upper-right port (P1) and emits the SWs at 15.2 GHz propagating to the left, as shown in Fig.~\ref{fig:circulator}(a). We observe the very weak transfer of energy to the Py stripe, so the SW propagates mainly in the top Co layer. The SWs of low intensity in the bottom Co layer result from direct magnetostatic coupling between Co layers. The power loss in the output (upper-left) port (P2) reaches 5.1 dB, while in the lower-left (P3) and lower-right (P4), we got high isolation, amounting 15.7 and 18.6 dB, respectively \footnote{Assuming no damping in the structure, the power loss in the port P2 reaches 0.3 dB, port P3 -- 13.7 dB, and port P4 -- 18 dB.}. The coupling direction is shown in Fig.~\ref{fig:circulator}(b). Here, the SW source is located in the upper-left port (P2). The SW is transferred to the Py stripe, and it reflects from the right edge of the stripe. After the reflection, the SW is coupled with the bottom Co layer, and, as a result, is transferred to it. The power loss in the output (lower-left) port (P3) reaches 7.7 dB, while in the lower-right (P4) and upper-right (P1), we got again high isolation of 19.9 and 32.9 dB, respectively \footnote{Assuming no damping in the structure, the power loss in the port P3 reaches 0.4 dB, port P4 -- 16.9 dB, and port P1 -- 24 dB.}.

The SW circulator can also be used as a SW diode. However, it benefits the mechanism of the redirection rather than the attenuation of a SW. Considering port P1 and P2 as the input/output ports, the transmission from port P1 to port P2 works as a forward direction and the transmission from port P2 to port P1 as a reverse direction. In that case, the difference in the SW energy in the forward and reverse direction equals 27.8 dB, thus even higher isolation properties than previously presented SW diode.

Figs.~\ref{fig:circulator} (c) and (d) show the power loss in the circulator as the function of NM layer thickness for the input port P1 and P2, respectively. We assumed $\alpha_{\text{Co}}=\alpha_{\text{Py}}=0$ to focus on the principle transmission properties of the system. The width of the Py stripe was set to the transmission length, which is plotted in Fig.~\ref{fig:transmission}(b). In the non-coupling case [Fig.~\ref{fig:circulator}(c)], the power loss in the target port P2 is decreasing, reaching almost no loss for about 10 nm. The power loss in port P4 is decreasing linearly while in the port P3, we see the oscillations. This is the result of the resonance in the Py stripe. This behavior is even more relevant in Fig.~\ref{fig:circulator}(d) representing the coupling case. The power loss in the target port P3 is oscillating in-counterphase towards the port P1. The points with large power loss in the port P3 correspond to the width $w$ of the Py stripe fulfilling the resonance condition $w = N \lambda/2$, where $\lambda=100$ nm. In the resonance, the SW is reflecting from the left side of the Py stripe, and it is coupled with the top Co layer. As a result, we observe the increase of the intensity of the SW in the port P1 and, simultaneously, decrease of the intensity in the port P3. Interestingly, the effect of negative power loss occurs in Fig.~\ref{fig:circulator}(d). It comes from the unwanted effect of the weak coupling between Co layers. For the thin NM layer, the coupling is significant enough to reach weak SW transmission from the top to bottom Co layer. In that case, we measured the SW energy in the range where we get the maximum value of the transmission. Moreover, the method of calculating the SW energy does not distinguish between the SW propagating in the left and right, which can fix this misleading effect. Also, because of the weak transmission between Co layers, the power loss can vary in dependence on the position of the antenna.

A slightly modified structure can be proposed for the realization of the three-port circulator for SWs. Its working principle is described in the Supplementary Materials.

\section{Discussion}

To sum up, we showed the effect of unidirectional magnetostatic coupling between the SW modes, which arises from the iDMI-induced nonreciprocity in the ultra-thin multilayer system. The modes related to each layer are strongly coupled in only one direction of the SW propagation in a broad GHz-range of frequency. In the opposite direction, within the same range of frequency, the SW modes propagate in only one layer. We proposed to exploit this effect for the realization of the magnonic devices with the sub-micrometer size. In the Py/NM/Co/Pt structure, limiting the Co/Pt stripe width to the length required to transfer the SW from the Py layer to the stripe, we arise the possibility to get the diode effect. In the forward direction, the SW propagates through the stripe area with small losses associated mainly with the Gilbert damping in Py, while in the reverse direction, the SW transmits to the Co stripe, in which the strong damping significantly reduces the SW intensity in the output. Importantly, the device works efficiently in a broad range of microwave frequencies. The SW diode can be further improved by opening the possibility to control the magnetization amplitude in the Co layer and thus becoming the SW transistor \cite{graczyk2020nonresonant, Nikitchenko2019, Duan2014, Padron-Hernandez2011, Khitun2009, Balinskiy2018, An2014}. Another proposed type of the magnonic device, which bases on the same effect, is the SW circulator. It uses the two extended Co layers with Pt inducing iDMI as the waveguides with the input/output ports and the Py stripe in between as a coupler. In each case, we get efficient SW transfer to the target port with the strongly suppressed signal at the other ports. Importantly, in the circulator, the isolation effect in the selected output ports is achieved without involving losses. Moreover, the SW circulator can also work as a diode. A diode and a circulator take an important place among the signal processing devices, thus demonstrated unidirectional coupling and proposed magnonic devices open new possibilities for further development of energy-efficient, miniaturized beyond-CMOS, magnonic logic components  \cite{au2012nanoscale,chumak2014magnon,cheng2016antiferromagnetic,cramer2018magnon}.

\section{Materials and Methods}

\subsection{Magnetization dynamics}

Magnetization dynamics in the systems under investigation are described by the Landau-Lifshitz-Gilbert equation:
\begin{equation}\label{eq:LL}
    \frac{\partial\textbf{M}}{\partial t} = -\gamma \mu_0 \textbf{M} \times \textbf{H}_{\text{eff}} + \frac{\alpha}{M_\text{S}} \textbf{M} \times \frac{\partial\textbf{M}}{\partial t},
\end{equation}
where $\textbf{M} = (m_x,m_y,m_z)$ is the magnetization vector, $\gamma$ is the gyromagnetic ratio, $\mu_0$ is the magnetic permeability of vacuum, and $H_{\text{eff}}$ is the effective magnetic field, which is given as follows:
\begin{equation}
    \textbf{H}_{\text{eff}} = H_0 \hat{z} + \frac{2A_{\text{ex}}}{\mu_0 M_S^2}\nabla^2\textbf{M} + \frac{2D}{\mu_0 M_S^2} \left(\hat{z} \times \frac{\partial \textbf{M}}{\partial x}\right) - \nabla \varphi,
\end{equation}
where $A_{\text{ex}}$ is the exchange stiffness constant, $D$ is the iDMI parameter, and $\varphi$ is the magnetic scalar potential fulfilling Maxwell equations in a magnetostatic approximation:
\begin{equation}\label{eq:pot}
    \nabla^2 \varphi = \nabla\cdot\textbf{M}.
\end{equation}

Eqs.~(\ref{eq:LL}) and (\ref{eq:pot}) are solved numerically in the linear approximation, i.e., assuming $m_x,m_y \ll m_z \approx M_S$, where $M_S$ is saturation magnetization, using the COMSOL Multiphysics. Frequency-domain simulations were carried out to calculate the SW dispersion relation in the system of coupled FM layers. Time-domain simulations were performed to demonstrate the functionality of the designed devices. A dynamic magnetic field is used to excite the system sinusoidally at the desired frequency. We use triangular mesh with a maximum element size of 1 nm inside the FM layers and a growth rate of 1.15 outside of the FM layers.

\subsection{Procedure for achieving the unidirectional coupling in a wide frequency range}

The effect of unidirectional coupling of SWs in a wide frequency range can be obtained according to the following procedure. We limit our approach to the Landau-Lifshitz equation consisting of the Zeeman, exchange, magnetostatic, and Dzyaloshinskii-Moriya terms [Eq.~(\ref{eq:LL})].

At first, we assume that the external magnetic field is uniform. Next, one should fulfill a condition, that
\begin{equation*}
    \begin{split}
 &\text{if }{M_{\text{S,FM1}} >(<)\,\, M_{\text{S,FM2}}}, \\
  &\text{then }   {A_{\text{ex,FM1}}/M_{\text{S,FM1}} >(<)\,\, A_{\text{ex,FM2}}/M_{\text{S,FM2}}}.
    \end{split}
\end{equation*}
It yields the non-crossing of the dispersion relation between non-interacting bilayers. If this condition is not fulfilled, we always will get crossing of the dispersion relations, and the coupling can be only asymmetric rather than unidirectional. Moreover, it is difficult to obtain the effect of coupling in a wide frequency range without fulfilling this condition. In the last step, the DMI parameter has to be fitted to get proper matching of dispersion relations for non-interacting layers. 

\section{Acknowledgments}

\begin{acknowledgments}
The study has received financial support from the National Science Center of Poland, projects no. \mbox{UMO-2018/30/Q/ST3/00416} and \mbox{UMO-2018/28/C/ST3/00052}. MM acknowledges funding from the Slovak Grant Agency APVV, no. \mbox{APVV-16-0068} (NanoSky).
\end{acknowledgments}

\bibliography{main} 

\end{document}


\title{SUPPLEMENTARY MATERIALS \\ 
Spin-wave diode and circulator based on unidirectional coupling}

\author{K.~Szulc$^{1}$ }
\author{P.~Graczyk$^{2}$ }
\author{M.~Mruczkiewicz$^{3}$ }
\author{G.~Gubbiotti$^{4}$ }
\author{M.~Krawczyk$^{1}$ }
\email{krawczyk@amu.edu.pl}
\affiliation{%
$^{1}$Faculty of Physics, Adam Mickiewicz University in Poznan, Uniwersytetu Pozna\'{n}skiego 2, 61-614 Pozna\'{n}, Poland 
}%
\affiliation{%
$^{2}$Institute of Molecular Physics, Polish Academy of Sciences, M. Smoluchowskiego 17, 60-179 Pozna\'{n}, Poland
}%
\affiliation{%
$^{3}$Centre of Excellence for Advanced Materials Application (CEMEA), Slovak Academy of Sciences, D\'{u}bravsk\'{a} cesta 9, 845 11 Bratislava, Slovakia 
}%
\affiliation{%
$^{4}$Istituto Officina dei Materiali del CNR (CNR-IOM), Sede Secondaria di Perugia, c/o Dipartimento di Fisica e Geologia, Universit\`{a} di Perugia, I-06123 Perugia, Italy
}%

\pacs{75.30.Ds, 75.40.Gb, 75.75.-c, 76.50.+g}

\maketitle

\section{Three-port spin-wave circulator}

\begin{figure*}[!h]
    \centering
    \includegraphics[width=\textwidth]{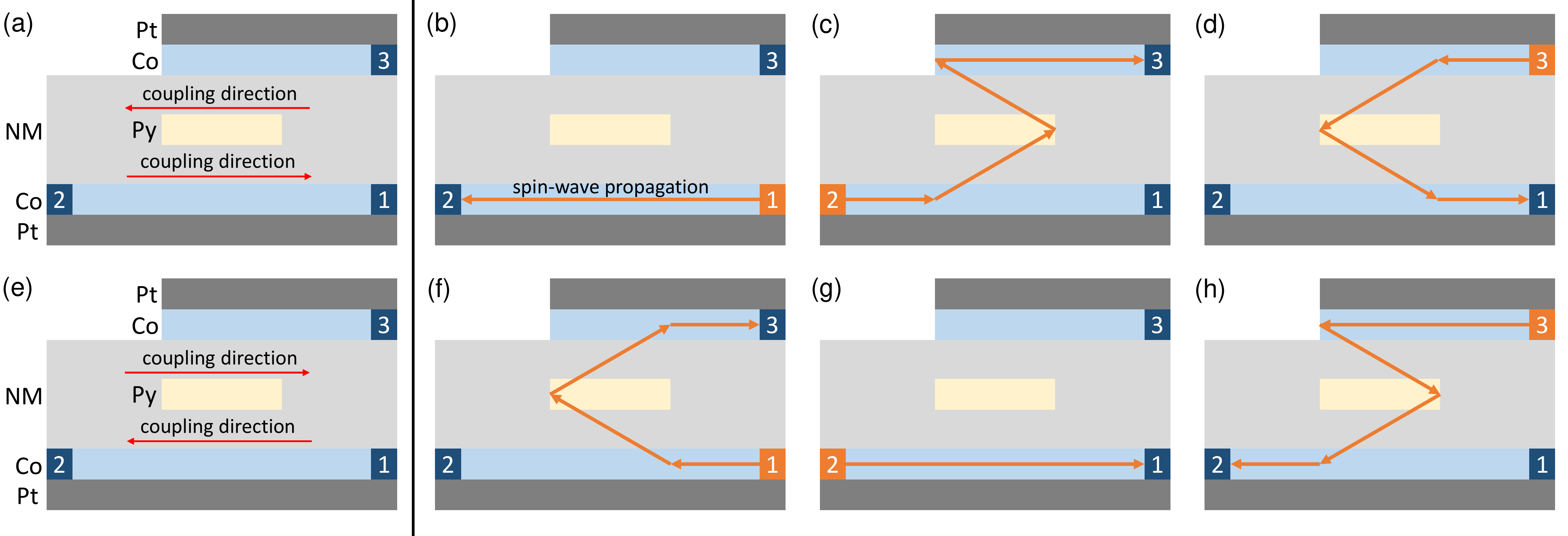}
    \caption{Three-port SW circulator in (a-d) easy-input and (e-h) easy-output port (3) configurations. We present the models (a, e) and the way of acting when the input of SWs is localized in (b, f) port (1), (c, g) port 2, and (d, h) port (3). The coupling direction determines the direction in which the SW will be transmitted from one layer to another. The orange arrows directing from one layer to another denote the SW transmission.}
    \label{fig:circulator3}
\end{figure*}

Along with the four-port circulator, we propose the three-port circulator with a slightly modified structure, as shown in Fig.~\ref{fig:circulator3}. As compared to the structure of the four-port circulator, here the upper Co/Pt layer length is reduced and consists now from one port. Furthermore, this layer has to fully cover the Py stripe to preserve the possibility of complete transfer of the SW between the layers. The three-port device does not have any symmetries, therefore, we have to take three cases into account independently. Moreover, the number of cases is doubled due to the reversal of the coupling directions. We can distinguish them by considering the efficiency of the port (3) localized in the upper Co layer. If the coupling direction between the upper Co layer and Py stripe is directed onto the port (3), we can consider it as an easy-output port because the SW coming from the Py stripe will be directed straight into the port (3) [Fig.~\ref{fig:circulator3}(f)]. On the other hand, with the opposite coupling direction, the SW coming from the port (3) will be transmitted directly to the Py stripe [Fig.~\ref{fig:circulator3}(d)], and the port (3) is an easy-input now. Fig.~\ref{fig:circulator3} shows that the wave propagating from the port (2) or (3) in the easy-input or easy-output configuration, respectively, have to reflect two times before reaching the output port [Figs.~\ref{fig:circulator3}(c) and (h)]. On the other hand, in the four-port circulator, only one reflection was needed. It means that the phenomenon in the three-port circulator requires very efficient reflections of SWs from the edges of the stripe.

\bibliography{main}